\documentclass[twocolumn,nofootinbib,floatfix,amsmath,amssymb,showpacs,superscriptaddress]{revtex4}

\usepackage{psfrag}
\usepackage{graphicx}
\usepackage{dcolumn}
\usepackage{bm}
\usepackage{color}

\usepackage{ulem}
\usepackage{color}

\begin{document}

\title{Colored-hadron distribution in hadron scattering in SU(2) lattice QCD}

\author{Toru T. Takahashi}
\affiliation{Gunma National College of Technology, Maebashi, Gunma
371-8530, Japan}
\author{Yoshiko Kanada-En'yo}
\affiliation{Department of Physics, Kyoto University, 
Sakyo, Kyoto 606-8502, Japan}

\date{\today}

\begin{abstract}
In color SU(2) lattice QCD, we investigate 
colored-diquark distributions in two-hadron scatterings 
by means of Bethe-Salpeter amplitudes on the lattice.
With colored-diquark operators in the Coulomb gauge,
we measure components of two colored diquarks
realized {as intermediate states} 
via one gluon exchange (OGE) processes in hadron scattering.
From the colored-diquark distributions, 
we estimate the dominant range of gluon (color) exchanges 
between closely located two hadrons.
We find that the colored-diquark components are enhanced at the short 
range ($<$0.2 fm)
and their tails show the single-exponential damping.
In order to distinguish the genuine colored-diquark components originating in the color exchange processes
from trivial colored two-quark components 
contained in two color-singlet hadrons 
as a result of 
simple transformation of hadronic basis,@
we repeat the analyses on the artificially constructed gauge fields,
where low- and high-momentum gluon components are decoupled and 
only restricted pair of quarks can share and exchange low-momentum gluons.
We observe qualitatively the same behaviors
and confirm that the short-range enhancement of the colored-diquark distributions
is the genuine
OGE-origin color excitation in hadron scattering.
\end{abstract}
\maketitle

\section{Introduction}

Hadron-hadron interactions play important roles in nuclear and hadron physics.
While the long- and the intermediate-range interactions are well
described by one-boson-exchange potentials (OBEP)~\cite{Yukawa:1935xg,PTPS39:1967,RMP39:1967,Machleidt:1987hj},
the short-range part of the hadronic interactions
has not been well clarified so far.
When hadrons are located closely each other,
the internal quark structures are of much importance
and interactions should be described in terms of quarks and gluons.
In recent lattice studies, 
the short-range interactions have been investigated based on QCD,
the fundamental theory of the strong interaction,
and especially the observed repulsive cores in the baryon interactions
on the lattice are found to be consistent with the quark-model based
interpretation~\cite{Oka:1980ax,Oka:1981ri,Oka:1981rj,Ishii:2006ec,Aoki:2012tk},
where the color-magnetic interactions via one gluon exchange (OGE) processes and Pauli blocking effects
among quarks play an essential role for the origin of repulsive cores.

In Ref.~\cite{Takahashi:2009ef}, we evaluated hadron-hadron interactions in color SU(2) QCD
by means of Bethe-Salpeter amplitudes on the lattice.
In the previous work, three different types of hadronic interactions were actually found. 
One is the middle-range universal attractive force,
which can be found in all the interaction channels and 
is independent from quark masses.
It is expected to be generated
by nonperturbative gluonic degrees of freedom.
Second is the short-range repulsive force,
for which the Pauli blocking effect
and the color-magnetic interaction among quarks are essential.
The last is the short-range attractive force, which is clearly seen 
in the four-flavor channel without the Pauli blocking effect. 
This short-range attraction has the strong quark-mass dependence and 
would be originated from gluon exchanges among quarks.
As just clarified, gluon exchange
is an important piece in hadron interactions besides boson exchanges.

Such gluon-exchange processes would be also important
from the view point of color exchanges between hadrons.
In a two-hadron scattering at a long distance, 
the color exchange does not occur and  
quarks in a color-singlet hadron are all confined. 
Only color-singlet objects, like mesons or glueballs, can be exchanged between hadrons.
At a short distance, however, 
colors can be exchanged via gluon exchanges between hadrons, corresponding to
the transition from two hadrons to  colored many-quark objects~\cite{Harvey:1980rva,Okiharu:2004ve,Okiharu:2004wy}. 
The intermediate state of the two colored objects  should change into two hadrons 
through recombination of quarks, 
or it gets back to the initial two-hadron channel with another color exchange.
The former process corresponds to the quark exchange between two 
hadrons via color exchanges.
Then, at the density scale where colors are actively exchanged among hadrons,
quarks would freely cross among hadrons
exhibiting effective quark deconfinement~\cite{Baym:1976yu,McLerran:2007qj,Abuki:2008nm,Miura:2008gd,McLerran:2008ua,Brauner:2009gu,Hands:2010gd,Fukushima:2015bda}.

\begin{figure}[h]
\begin{center}
\includegraphics[scale=0.35]{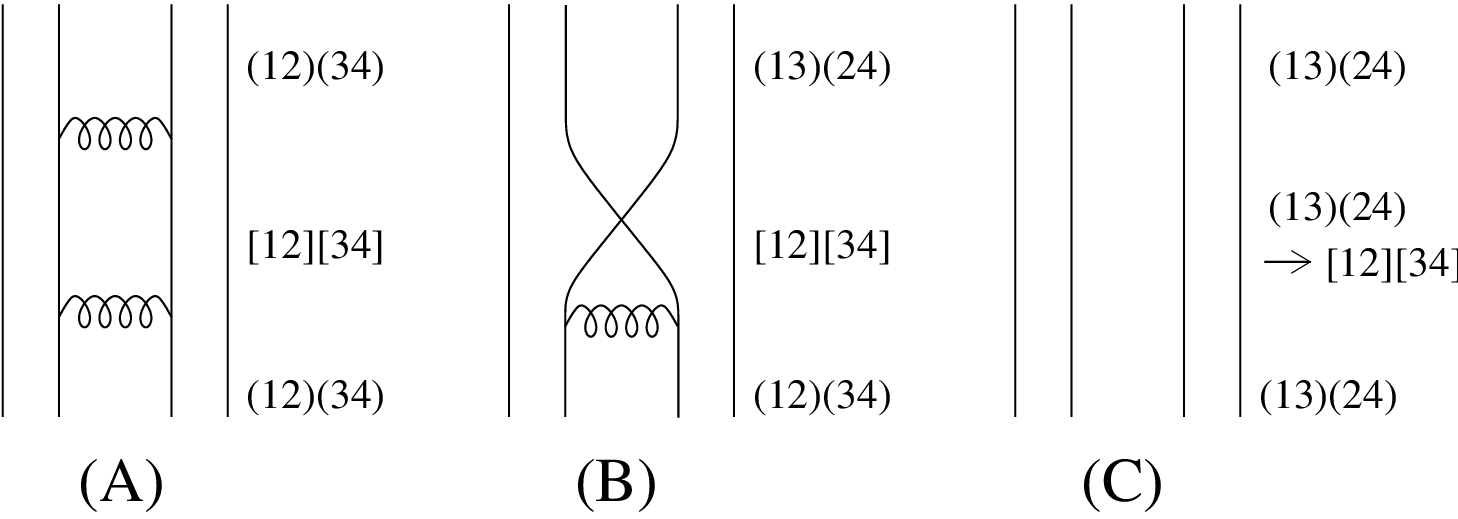}
\end{center}
\caption{
(A)Incoming (12)(34) diquarks exchange colors
and form outgoing (12)(34) diquarks.
(B)Incoming (12)(34) diquarks exchange colors 
and form outgoing (13)(24) diquarks, rearranging quark pairs.
(C)Incoming and outgoing diquarks are (13)(24),
which however has non-zero overlap with [12][34] operator
(Trivial colored two-quark component).
\label{Fig.COLexc}}
\end{figure}

In the case of color SU(2) QCD, which we employ in the present analysis,
baryons are color-singlet diquarks.
Let us consider (12)(34) incoming two diquarks,
where participating quarks have flavor 1, 2, 3 and 4,
and 1,2-  and 3,4- quark pairs form color-singlet diquarks,
(12) and (34).
If one gluon is exchanged between 1- and 3-quarks,
it results in color exchange between 12- and 34-diquarks,
and 12-  and 34-diquarks get colored, forming a [12][34] two-diquark system.
(Here [ij] denotes a color triplet ij-diquark.)
Some possible gluon exchange processes are illustrated in Fig.~\ref{Fig.COLexc}(A)(B).
These gluon exchange processes, including quark recombination processes,
can be detected by probing [12][34] colored-diquark components (distribution),
since any gluon exchange processes go through a [12][34] two-diquark state.
However, there exist trivial colored two-quark components.
(13)(24) two-diquark wavefunction contains non-zero [12][34] component
as easily proved when we rewrite it 
in the quark basis of 12- and 34- diquarks (Fig.~\ref{Fig.COLexc}(C)).
We should distinguish the genuine OGE-origin colored-diquark components from such trivial contributions.

In this article, 
we investigate gluon (color) exchanges between closely located two hadrons
with the SU(2) lattice QCD,
by looking at the colored-diquark distributions between two scattering hadrons.
Especially we concentrate on the color-exchange range between two hadrons
to clarify the origin of the short-range attraction in the four-flavor channel
free from the Pauli blocking effect.
In Sec.\ref{Sec.Formulation}, the detail of our formulation is presented.
The results are shown in Sec.\ref{Sec.Lattice}, and Sec.\ref{Sec.Add} is devoted to the additional analysis to confirm the results.
Results are discussed in Sec.\ref{Sec.Discussions}
and our present study is summarized in Sec.\ref{Sec.Summary}.

\section{Formulation}
\label{Sec.Formulation}

We follow the strategy proposed by CP-PACS group~\cite{Aoki:2005uf,Aoki:2009ji}, 
where ``wavefunctions'' are defined as Bethe-Salpeter amplitudes
measured with quark operators.
We define local diquark operators $\phi_{\{i,j\},\Gamma}^c({\bm x},t)$ as
\begin{eqnarray}
\phi_{(i,j),\Gamma}({\bm x},t)
\equiv
\frac{1}{\sqrt{2}}
\varepsilon_{ab}
\Gamma_{\alpha\beta}
q_i^{a\alpha}({\bm x},t)
q_j^{b\beta}({\bm x},t).
\end{eqnarray}
for color-singlet diquarks, and as
\begin{eqnarray}
\phi_{[i,j],\Gamma}^c({\bm x},t)
\equiv
\frac{1}{\sqrt{6}}
T_{ab}^c
\Gamma^{\alpha\beta}
q_i^{a\alpha}({\bm x},t)
q_j^{b\beta}({\bm x},t).
\end{eqnarray}
for color-triplet diquarks.
The subscripts $i$ and $j$ denote quark flavors
and $(i,j)$ and $[i,j]$ show that $i,j$-quark pair
belongs to a color-singlet and color-triplet state, respectively.

$q_i^{a\alpha}({\bm x},t)$ is a quark operator located at $({\bm x},t)$
that has the color index $a$, the spinor index $\alpha$ and the flavor
index $i$.
$\varepsilon_{ab}$ is an antisymmetric tensor 
with $\varepsilon_{12}=1$
and $T_{ab}^c$
is a tensor that projects the diquark onto a color-triplet state (vector representation).
$\Gamma$ set to $C\gamma_5$ ($C\gamma_\mu$)
makes a $J=0$ ($J=1$) diquark state.

In order to measure the relative wavefunction in diquark-diquark scattering,
we compute two-diquark correlators $W_{\{i,j\}\{k,l\}}^\Gamma({\bm R},t)$ defined as
\begin{eqnarray}
&&W_{(i,j)(k,l)}^\Gamma({\bm R},t)
\equiv\nonumber\\
&&\sum_{\bm x}
\langle
\phi_{(i,j),\Gamma}({\bm x},t) \phi_{(k,l),\Gamma}({\bm x}+{\bm R},t)\nonumber\\
&&\hspace{8em}
\phi_{(i,j),\Gamma}^\dagger({\bm 0},0) \phi_{(k,l),\Gamma}^\dagger({\bm 0},0)
\rangle
\end{eqnarray}
and
\begin{eqnarray}
&&W_{[i,j][k,l]}^\Gamma({\bm R},t)
\equiv\nonumber\\
&&\sum_{{\bm x},c,c'}
\langle
\phi_{[i,j],\Gamma}^c({\bm x},t) \phi_{[k,l],\Gamma}^c({\bm x}+{\bm R},t)\nonumber\\
&&\hspace{8em}
\phi_{[i,j],\Gamma}^{c'\dagger}({\bm 0},0) \phi_{[k,l],\Gamma}^{c'\dagger}({\bm 0},0)
\rangle.
\end{eqnarray}
The summation over $c$ and $c'$ is taken to make the system totally color singlet.
${\bm R}$ represents the relative coordinate between two scattering hadrons.
Two-diquark operators are all projected to totally spin-0 states
through the contraction over $\mu$ in $\Gamma$.
We employ wall-type operators for sources, 
while we use point-type operators for sinks.
We hereby denote
$W_{\{i,j\}\{k,l\}}^{C\gamma_5}({\bm R},t)$
and
$W_{\{i,j\}\{k,l\}}^{C\gamma_\mu}({\bm R},t)$
as
\begin{eqnarray}
\Phi_{\{i,j\}\{k,l\}{\rm S}}({\bm R},t)
\equiv
W_{\{i,j\}\{k,l\}}^{C\gamma_5}({\bm R},t)
\end{eqnarray}
and
\begin{eqnarray}
\Phi_{\{i,j\}\{k,l\}{\rm A}}({\bm R},t)
\equiv
W_{\{i,j\}\{k,l\}}^{C\gamma_\mu}({\bm R},t),
\end{eqnarray}
throughout this paper.
$\Phi_{(i,j)(k,l){\rm S,A}}({\bm R},t)$ 
and
$\Phi_{\{i,j\}\{k,l\}{\rm S,A}}({\bm R},t)$ 
give color-singlet and color-triplet diquark distributions
in hadron scattering.
Especially, 
color excitation mode in hadron scattering
can be probed by the color-triplet diquark distribution.

The internal wave functions of isolated diquarks
probed by local quark operators are defined as 
\begin{eqnarray}
&&
\Phi_{(i,j)S,A}({\bm R},t)
\nonumber \\
&\equiv&
\sum_{{\bm x}}
\frac{1}{\sqrt{2}}
\varepsilon_{ab}
\Gamma_{\alpha\beta}
\langle
q_i^{a\alpha}({\bm x}+{\bm R},t)
q_j^{b\beta}({\bm x},t)\nonumber \\
&&\hspace{12em}
\phi_{(i,j)S,A}^\dagger({\bm 0},0)
\rangle
.
\end{eqnarray}
$\Gamma$ is chosen as $C\gamma_5$ and $C\gamma_\mu$ 
for scalar and axial vector diquarks, $\Phi_{(i,j)S}$ and $\Phi_{(i,j)A}$, respectively. ${\bm R}$ represents the relative coordinate between two quarks in the diquark.

We concentrate on S-wave scattering states of
two scalar diquarks ($\Gamma = C\gamma_5$),
which is the ground state in $l=0$ channel,
and we project wavefunctions $\Phi({\bm R})$ 
onto $A_1^+$-wavefunction $\Phi(R)\equiv \Phi(|{\bm R}|)$,
which has overlap with $l=0$ states,
by summing up $\Phi({\bm R})$  in terms of corresponding discrete rotations.
We here neglect the contributions from $l\geq 4$ scattering states,
since such contributions can be dropped by taking large Euclidean-time separation $t$.

All the simulations are performed in SU(2) quenched QCD
with the standard plaquette gauge action and the Wilson quark action.
The lattice size is $24^3\times 64$ at $\beta = 2.45$,
whose lattice spacing is about 0.1 fm
if we assume $\sqrt{\sigma}$ is 440 MeV~\cite{Fingberg:1992ju,Stack:1994wm}.
We employ four different Hopping parameters 
$\kappa$ = 0.1450, 0.1475, 0.1500, 0.1513 for quarks.

\section{Lattice QCD results}
\label{Sec.Lattice}

\subsection{Hadron masses}
\label{Sec.LatticeHadron}

\begin{figure}[h]
\begin{center}
\includegraphics[scale=0.28]{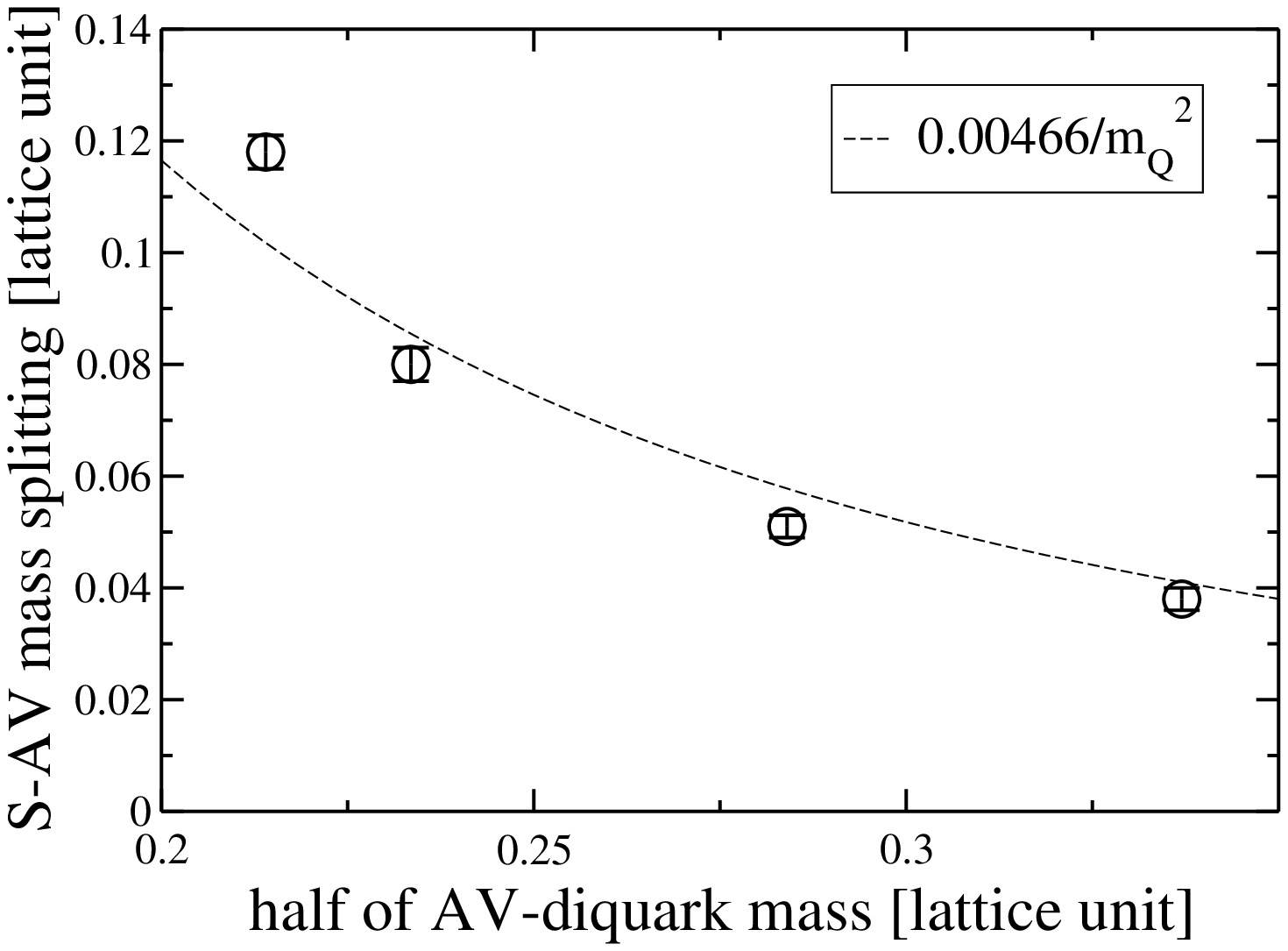}
\end{center}
\caption{
Scalar-axialvector mass splitting is plotted 
as a function of axialvector-diquark mass.
The dotted line is a fit function $Cm_{\rm Q}^{-2}$.
\label{Fig.pirho}}
\end{figure}

We show the ground-state diquark masses 
for each quantum number in Table~\ref{Tab.hadronicmasses}.
In Fig.~\ref{Fig.pirho},
scalar-axialvector mass splittings $\Delta m$ are plotted
as a function of the half of the axialvector diquark masses,
which can be regarded as ``constituent'' quark masses $m_Q$.
The dotted line denotes the fit function
$\Delta m = C/m_Q^2$.
Most of the data can be well reproduced by
$\Delta m = C/m_Q^2$,
although $\Delta m$ at the lightest quark mass
($m_\pi\simeq 600$ MeV)
deviates from the fit function,
which implies a naive quark model picture
may not be valid any longer in this quark-mass region.

\begin{table}[h]
\begin{tabular}{llllll}
\hline
$\kappa$ & Scalar & Axialvector & Pseudoscalar & Vector & $\Delta m$\\ \hline\hline
0.1450            & 0.636(2) & 0.674(3) & 0.893( 19) & 0.897(13) & 0.038(2)
\\
0.1475            & 0.517(2) & 0.568(4) & 0.811( 28) & 0.805(17) & 0.051(2)
\\
0.1500            & 0.387(3) & 0.467(7) & 0.768( 54) & 0.714(24) & 0.080(3)
\\
0.1513            & 0.310(3) & 0.428(1) & 0.717(116) & 0.673(30) & 0.118(3)
\\ \hline\hline
\end{tabular}
\caption{\label{Tab.hadronicmasses}
All the hadronic masses are listed.
$\Delta m$ represents the scalar-axialvector diquark mass splitting.
}
\end{table}

\subsection{Wavefunctions}

Wavefunctions are measured from two-diquark operators
for
$(1,2)(3,4)S$,    $[1,2][3,4]A$
$(1,3)(2,4)S$,    $(1,3)(2,4)A$
$[1,3][2,4]S$ and $[1,3][2,4]A$
channels.
Here, $(i,j)(k,l)$ ($[i,j][k,l]$) mean
that
$ij$- and $kl$-diquarks belong to
color-singlet (color-triplet) states.
Such 4-quark operators are all projected to totally
color-singlet and $J=0$ operators.
Wavefunction-analyses with $[i,j][k,l]$ operators
allow us to probe the color-excitation modes in hadron scatterings.
We note that $(1,2)(3,4)A$ and $[1,2][3,4]S$ scattering states
are prohibited,
since the 12-diquark can be either $(1,2)S$ or $[1,2]A$ state
due to the ``iso-spin conservation'' for 1- and 2-quarks.
(The initial state is in all the cases $(1,2)(3,4)S$ scattering state.)

\begin{figure}[h]
\begin{center}
\includegraphics[scale=0.3]{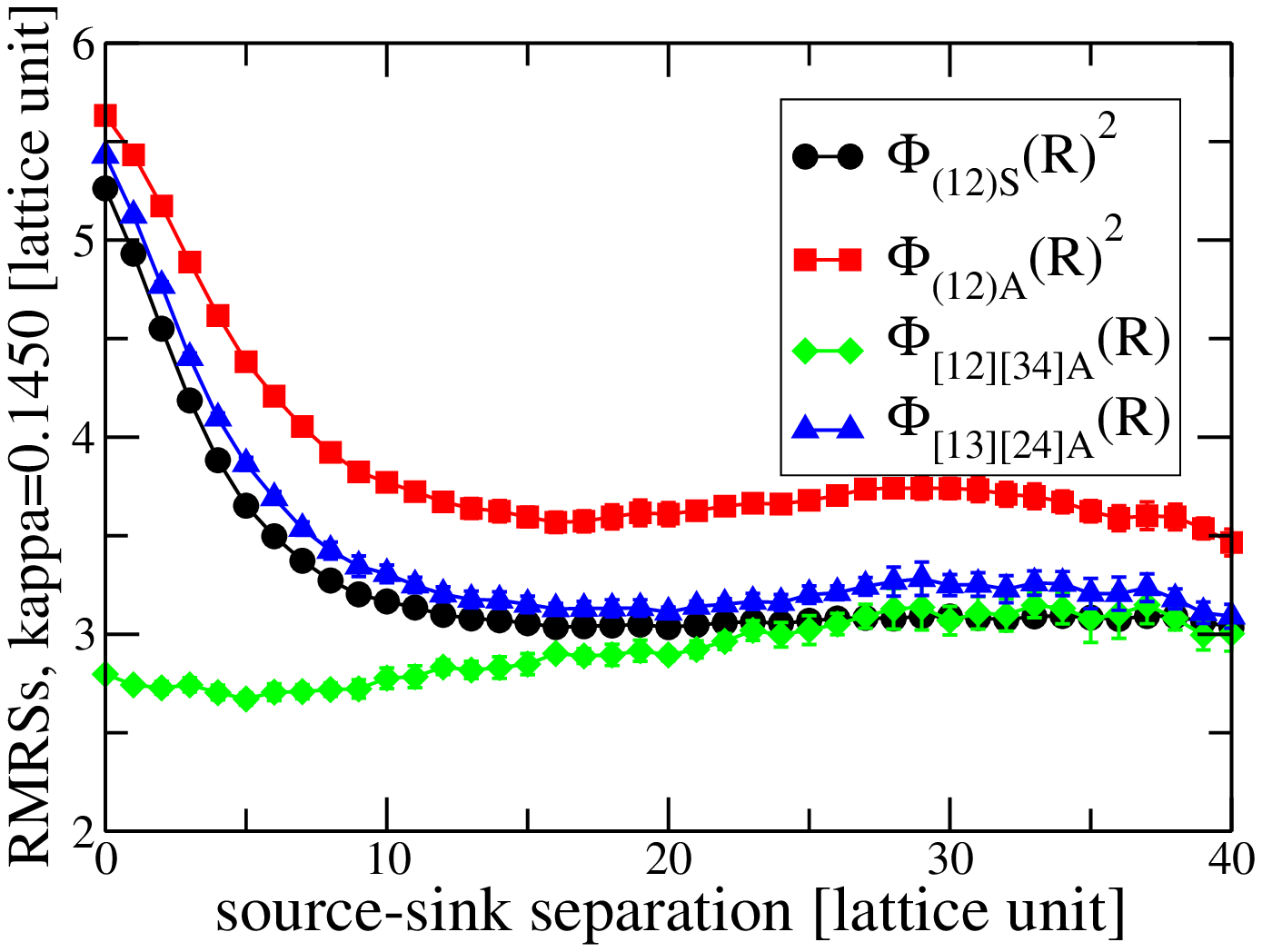}
\end{center}
\caption{
\label{Fig.RMRS}
The RMSRs obtained from wavefunctions $\Phi(R,t)$ 
at $\kappa=0.1450$ are plotted 
as a function of a source-sink separation $t$.
They are all normalized with $0\leq R \leq 7$.
}
\end{figure}

We here mention
the extraction of the wavefunctions $\Phi(R)$ using the Euclidean correlators.
In Fig.~\ref{Fig.RMRS},
the RMSRs (root mean square radii) obtained from the wavefunctions $\Phi(R,t)$ 
are plotted as functions of a source-sink separation $t$.
The RMSRs are computed for 
$[1,2][3,4]A$, $[1,3][2,4]S$ and $[1,3][2,4]A$ channels,
in which wavefunctions show exponential damping in the asymptotic
region and RMSRs can be well defined.
The time-dependent wavefunctions $\Phi(R,t)$ are all normalized with $0\leq R \leq 7$, since  at $R\sim 7$ they are small enough and would be safe from finite size effects.
The RMSRs evaluated from the internal wavefunctions of the scalar- and axialvector-diquarks, $\Phi_{(12)S}$ and $\Phi_{(12)A}$ are also displayed.
They first depend on $t$ due to excited-state contaminations,
but finally show plateaus at large $t$ region, which implies ground-state dominance.
All the wavefunctions $\Phi(R)$ are determined at $t=25$; $\Phi(R)\equiv\Phi(R,25)$.

\begin{figure}[h]
\begin{center}
\includegraphics[scale=0.6]{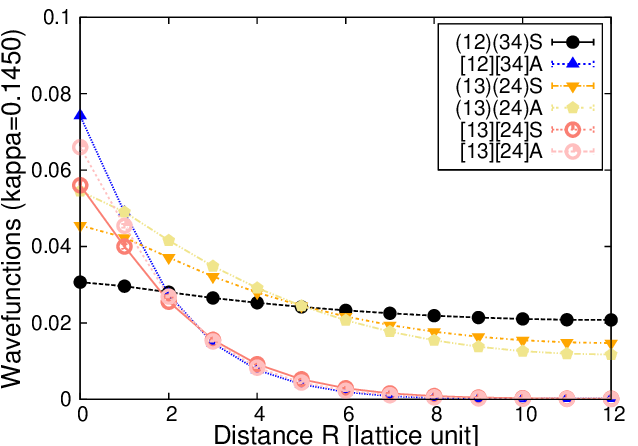}
\includegraphics[scale=0.6]{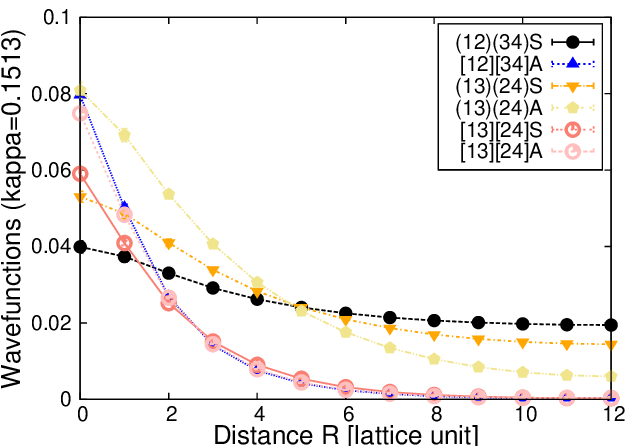}
\end{center}
\caption{
All the wavefunctions
are plotted as functions
of distance $R$ for $\kappa=$0.1450 and 0.1513.
They are normalized within $0\leq R \leq 7$.
\label{Fig.WFs}}
\end{figure}

\begin{figure}[h]
\begin{center}
\includegraphics[scale=0.6]{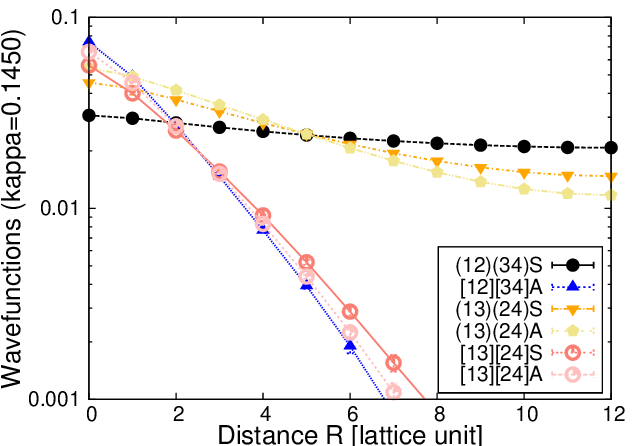}
\includegraphics[scale=0.6]{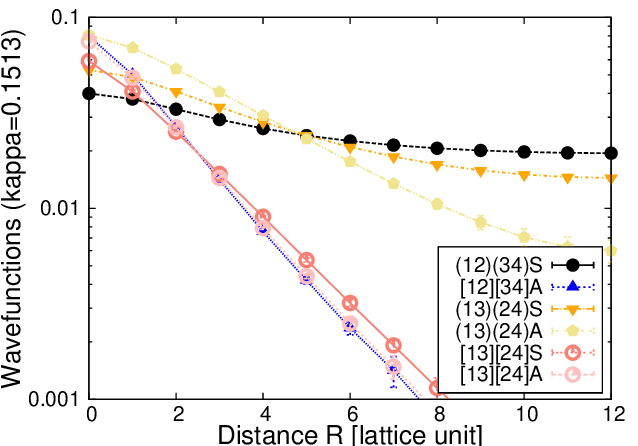}
\end{center}
\caption{
All the wavefunctions
are logarithmically
plotted as functions of distance $R$ for $\kappa=$0.1450 and 0.1513.
They are normalized within $0\leq R \leq 7$.
\label{Fig.WFs2}}
\end{figure}

In Fig.\ref{Fig.WFs}, all the wavefunctions 
$\Phi(R)$ are displayed as functions of the
relative coordinate $R$.
Their logarithmic plots are shown in Fig.\ref{Fig.WFs2}.
In $(1,2)(3,4)S$, $(1,3)(2,4)S$ and $(1,3)(2,4)A$ channels,
wavefunctions remain finite at the asymptotic region,
which shows that 
rearrangement of quarks is realized in hadron scatterings:
Only a 4-quark state with $(1,2)(3,4)S$ is created at $t=0$ (source point),
and after (Euclidean) time evolution, quarks are exchanged among hadrons
and other possible channels appear at the late Euclidean time.
On the other hand, wavefunctions measured by colored operators
($[1,2][3,4]A$, $[1,3][2,4]S$ and $[1,3][2,4]A$ channels)
exponentially damp in the asymptotic region.
These observations imply that
colored diquarks cannot appear as asymptotic states,
and are bounded at the short-distance region in hadron scatterings
due to the color confinement.
Such bounded colored diquarks form so-called ``multi-quark'' states.

In this paper, we especially concentrate on $[1,2][3,4]A$ wavefunctions,
which has no overlap with and is completely independent from 
incoming $(1,2)(3,4)S$ state.
In the channel,
each diquark operator belongs to a color-triplet state
and can be a probe for color excitations in hadron scatterings.

\subsection{Colored Wavefunctions}

\begin{figure}[h]
\begin{center}
\includegraphics[scale=0.6]{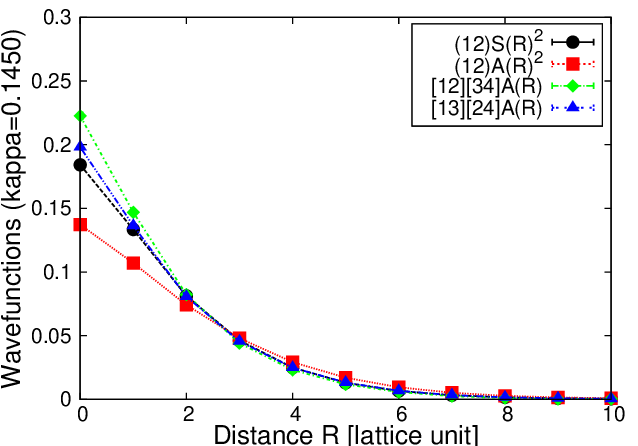}
\includegraphics[scale=0.6]{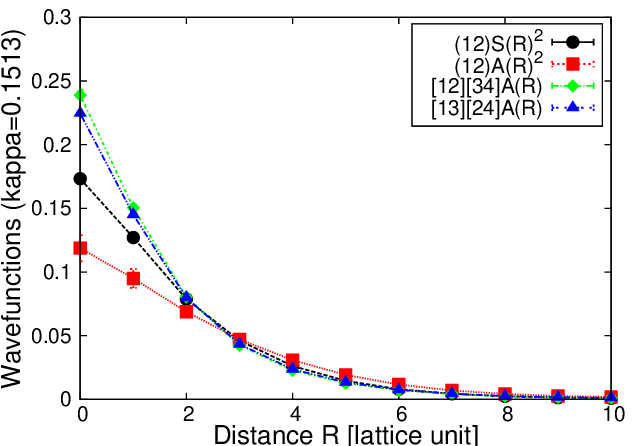}
\end{center}
\caption{
The wavefunctions measured by colored operators
as well as the squared wavefunctions of single hadrons,
which are normalized within $0\leq R \leq 7$, are plotted.
\label{Fig.WFs3}}
\end{figure}

In Fig.~\ref{Fig.WFs3}, we again show the wavefunctions 
$\Phi_{[12][34]A}(R)$ and $\Phi_{[13][24]A}(R)$
from BS amplitudes
as well as the squared internal wavefunctions, $\Phi_{(12)S}(R)^2$ and
$\Phi_{(12)A}(R)^2$ for scalar- and axialvector diquarks.

All the wavefunctions are normalized within $0\leq R \leq 7$
to avoid possible finite-size effects at the boundary.
As the quark mass decreases,
the internal wavefunction of the axialvector diquark gets broader,
while that of the scalar diquark shows little quark-mass dependence.
The colored-diquark distributions,
$\Phi_{[12][34]A}(R)$ and $\Phi_{[13][24]A}(R)$,
are remarkably enhanced at the short range for the larger $\kappa$,
which implies color exchange among quarks
is much more active for the lighter quark mass.
This is consistent with the scenario 
 proposed in Ref.~\cite{Takahashi:2009ef}:
The short-range attraction shows strong quark-mass dependence 
and may come from the 
color-magnetic interactions (color exchange).

We should note here that such color contributions evaluated from
$\Phi_{[12][34]A}(R)$ and $\Phi_{[13][24]A}(R)$
do not always indicate
the genuine OGE-origin colored-diquark components
(genuine color excitations)
but also include trivial colored two-quark contributions.
It comes from the fact that 
all the colored-diquark components can be always expressed
in terms of color-singlet degrees of freedom.
For example,
$\Phi_{[12][34]A}(R)$,
extracted by two colored-diquark fields $q_1({\bm 0})q_2({\bm 0})$ 
and $q_3({\bm R})q_4({\bm R})$ with separation $R$,
has non-zero overlap with
color-singlet diquark scattering wavefunction
$|(13)S(24)S\rangle$,
which consists of two color-singlet diquarks $|(13)S\rangle$ and $|(24)S\rangle$.
Namely, taking into account that $T_{ab}^cT_{a'b'}^c$ can be transformed by 
linear combination of 
$\varepsilon_{aa'}\varepsilon_{bb'}$ and $\varepsilon_{ab'}\varepsilon_{ba'}$,
the overlap is written as
\begin{eqnarray}
&&\sum_{c}
 \langle {\rm vac} | 
\phi_{[12]A}^c({\bm 0}) \phi_{[34]A}^c({\bm R})
|
(13)S(24)S\rangle 
\nonumber \\
&\sim&
T_{ab}^cT_{a'b'}^c  \langle {\rm vac} | 
q_1^a({\bm 0})q_2^b({\bm 0}) q_3^{a'}({\bm R})q_4^{b'}({\bm R})
|(13)S\rangle|(24)S\rangle
\nonumber \\
&=&
T_{ab}^cT_{a'b'}^c  
 \langle {\rm vac} | 
q_1^a({\bm 0})q_3^{a'}({\bm R})|(13)S\rangle
 \langle {\rm vac} | 
q_2^b({\bm 0}) q_4^{b'}({\bm R})
|(24)S\rangle
\nonumber \\
&\sim&
\varepsilon_{aa'}\varepsilon_{bb'}
 \langle {\rm vac} | 
q_1^a({\bm 0})q_3^{a'}({\bm R})|(13)S\rangle
 \langle {\rm vac} | 
q_2^b({\bm 0}) q_4^{b'}({\bm R})
|(24)S\rangle
\nonumber \\
&\sim&
\Phi_{(13)S}(R)\Phi_{(24)S}(R).
\end{eqnarray}
Here we omit some normalization factors and consider the extreme case that
the internal hadron structure does not change in the scattering.
In such cases, the $R$-dependence of $\Phi_{[12][34]A}(R)$
is expressed in terms of diquark (internal) wavefunctions.
If $\Phi_{[12][34]A}(R)$ is reflecting only
the quark-rearranged color-singlet scattering waves $|(13)(24)S\rangle$
or $|(13)(24)A\rangle$,
its $R$-dependence will resemble those of
squared internal wavefunctions,
$\Phi_{(12)S}(R)^2$ and $\Phi_{(12)A}(R)^2$,
of color-singlet diquarks.

\begin{figure}[h]
\begin{center}
\includegraphics[scale=0.6]{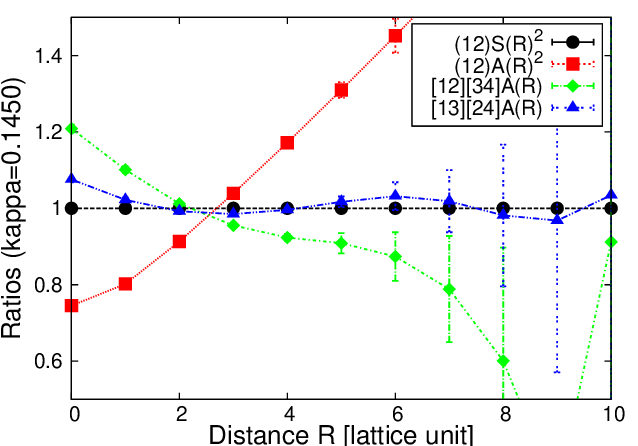}
\includegraphics[scale=0.6]{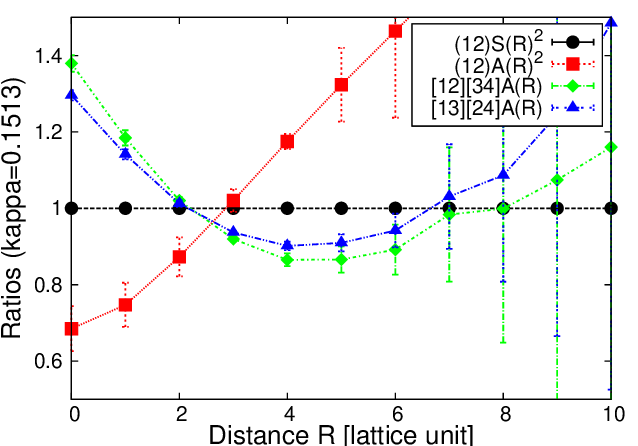}
\end{center}
\caption{
Ratios between the wavefunctions from colored operators
and the squared wavefunction of the scalar diquark hadron,
which are normalized within $0\leq R \leq 7$, are plotted.
\label{Fig.RTs}}
\end{figure}

In order to compare colored-diquark distributions and
color-singlet diquark (internal) wavefunctions,
in terms of $R$-dependence,
we show wavefunctions normalized by
$\Phi_{(12)S}(R)^2$ in Fig.~\ref{Fig.RTs}.
If
there is no genuine colored-diquark component and
$\Phi_{[12][34]A}(R)$
simply reflects trivial colored two-quark contributions,
the colored-diquark distribution functions
$\Phi_{[12][34]A}(R)$ and $\Phi_{[13][24]A}(R)$
will show $R$ dependences similar to
$\Phi_{(12)S}(R)^2$ and $\Phi_{(12)A}(R)^2$
(possible two color-singlet degrees of freedom).
In actuality, the colored-diquark distributions $\Phi_{[12][34]A}(R)$ and $\Phi_{[13][24]A}(R)$
at $R< 2$ lie above those two lines.
Especially $\Phi_{[12][34]A}(R)$ clearly lies above $\Phi_{(12)S}(R)^2$,
and this tendency is stronger at lighter quark mass region.
Then, enhanced color excitations are expected at $R< 2$.
In the case of $\Phi_{[13][24]A}(R)$, it is relatively smaller.
The reason is the asymmetry between $(12)(34)S$ and $(13)(24)S$
asymptotic states.
At the source point, only $(12)(34)S$ state is created
and this leads to the relative suppression of $(13)(24)S$ waves.
While $\Phi_{[13][24]A}(R)$ would surely contain colored-diquark
contribution,
it is smaller because of the suppression of $(13)(24)S$ incoming waves,
and it largely reflects the trivial colored two-quark components from the
(relatively larger) incoming and outgoing $(12)(34)S$ state.

\section{Additional analyses}
\label{Sec.Add}
\subsection{Isolating color excitations}
\label{Sec.Isolation}

QCD interactions may be classified into two categories.
One is highly nonperturbative dynamics,
which leads to the chiral-symmetry breaking, color confinement and etc,
and the other is perturbative contributions 
represented by one gluon exchange (OGE) interactions among quarks.
These two distinct phenomena originate from difference energy scales.
A method that restricts energy scale of gluons
via the Fourier transformation was proposed and tested~\cite{Yamamoto:2008am,Yamamoto:2008ze}.

One of the principal origins of color excitations
would be high-energy gluon exchanges (one gluon exchange) 
at the short-distance regions.
When a gluon is exchanged between 1- and 3-quarks in the (12)(34)S hadronic state, a [12][34]A state would be formed as an intermediate state.
As discussed before,
the [12][34]A component measured by colored-diquark operators cannot be fully distinguished from (13)(24)S or (14)(23)S hadronic states, since (13)(24)S and (14)(23)S states can be asymptotic states and are mixed in hadronic scattering to contribute to the colored-diquark components.
In order to make (13)(24)S and (14)(23)S asymptotic states absent from the system and isolate color excitations,
we make high-momentum gluons be common among all the quarks,
and let low-momentum gluons be independent.
If low-momentum gluons are made common between 1,2 and 3,4 quarks,
only the (12)(34)S asymptotic state survives but asymptotic states
of other channels (13)(24)S or (14)(23)S are prohibited,   
since confining force always acts only between 1,2 and 3,4 quarks.
(Not between 1,3 and 2,4 quarks.)

After this gluon-manipulation, 
the asymptotic region of the system is fulfilled only by
$(12)(34)S$ diquark scattering waves and 
$\Phi_{[12][34]A}(R)$ does not contain contaminations from
$(13)(24)S$ and $(14)(23)S$ scattering wavefunctions.

With this aim, 
we generate three different series of SU(2) link variables (gluon fields),
$U_\mu ({\bm x},t;C)$, $U_\mu ({\bm x},t;C')$ and $U_\mu ({\bm x},t;C'')$.
We perform 3-dim Fourier transformation~\cite{Yamamoto:2009na} 
for link variables $U_\mu ({\bm x},t)$ in the Landau gauge
and extract $U_\mu ({\bm p},t)$ in momentum space:
\begin{equation}
U_\mu ({\bm p},t)
=
\sum_{\bm x}
U_\mu ({\bm x},t)
e^{i{\bm p}\cdot{\bm x}}
\end{equation}
We divide link variables into low- and high-momentum contributions,
\begin{equation}
\widetilde{U}^{\rm L}_\mu ({\bm p},t)=
U_\mu ({\bm p},t)\ (|{\bm p}|\leq\Lambda) \\
\end{equation}
\begin{equation}
\widetilde{U}^{\rm H}_\mu ({\bm p},t)=
U_\mu ({\bm p},t)\ (|{\bm p}|\geq\Lambda) \\
\end{equation}
and construct mixed link variables as
\begin{eqnarray}
\widetilde{U}_\mu ({\bm p},t;C_A)
&=&
\widetilde{U}^{\rm L}_\mu ({\bm p},t;C')
+
\widetilde{U}^{\rm H}_\mu ({\bm p},t;C)
\\
\widetilde{U}_\mu ({\bm p},t;C_B)
&=&
\widetilde{U}^{\rm L}_\mu ({\bm p},t;C'')
+
\widetilde{U}^{\rm H}_\mu ({\bm p},t;C)
\end{eqnarray}

With
link variables reconverted into the coordinate representation,
\begin{equation}
\widetilde{U}_\mu ({\bm x},t)\equiv
\frac{1}{V}\sum_{\bm p}\widetilde{U}_\mu ({\bm p},t)e^{-i{\bm p}\cdot{\bm x}},
\end{equation}
which no longer belong to SU(2),
new SU(2) link variables in the coordinate space
$U_\mu ({\bm x},t)$
are reconstructed so that the distance,
\begin{equation}
{\rm Tr} \ 
(U_\mu ({\bm x},t)-\widetilde{U}_\mu ({\bm x},t))
(U_\mu ({\bm x},t)-\widetilde{U}_\mu ({\bm x},t))^\dagger,
\end{equation}
is minimized.
The infrared(IR) cut $\Lambda$ 
is set to $\Lambda = 5$ in lattice unit (about 1 GeV in the physical unit).
1,2-quarks' propagators are computed
with thus constructed link variables $U_\mu ({\bm x},t;C_A)$,
and for 3,4-quarks' propagators we use $U_\mu ({\bm x},t;C_B)$.
Then, $U_\mu ({\bm x},t;C_A)$ and $U_\mu ({\bm x},t;C_B)$,
give independent low-energy QCD dynamics to 1,2-quarks and
3,4-quarks but high-momentum dynamics are common in all the quarks.
Hadrons interacting with
$U_\mu ({\bm x},t;C_A)$ and $U_\mu ({\bm x},t;C_B)$
have nothing to do with each other in the asymptotic region,
but can exchange only high-momentum gluons in the short-distance region.
Using this nature, we may single out color exchanges in the
short-distance region;
we do not suffer from the trivial colored two-quark state coming from 
transformation of hadronic basis.
(Asymptotic two-hadron states in the (13)(24) and (14)(23) channels are
prohibited since $U_\mu ({\bm x},t;C_A)$ and $U_\mu ({\bm x},t;C_B)$ give
different confining forces.)
Note that we also performed Fourier transformation in the Coulomb gauge
and obtained qualitatively the same results.

\subsection{Isolated color excitations}
\label{Sec.Isolated}

\begin{figure}[h]
\begin{center}
\includegraphics[scale=0.6]{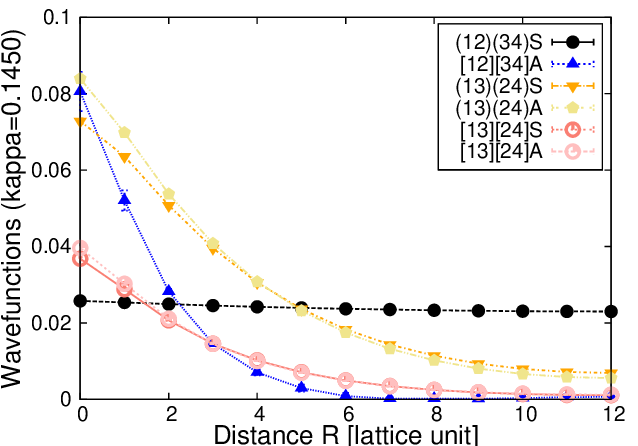}
\includegraphics[scale=0.6]{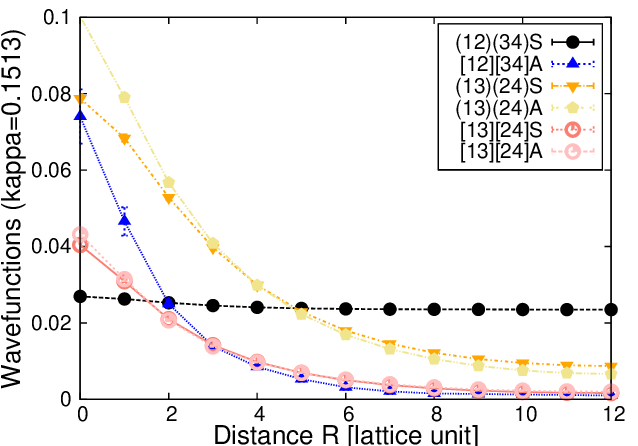}
\end{center}
\caption{
All the wavefunctions,
which are obtained without IR-gluon exchanges,
are
plotted as functions of distance $R$ for $\kappa=$0.1450 and 0.1513.
They are normalized within $0\leq R \leq 7$.
\label{Fig.WFsFFT}}
\end{figure}

\begin{figure}[h]
\begin{center}
\includegraphics[scale=0.6]{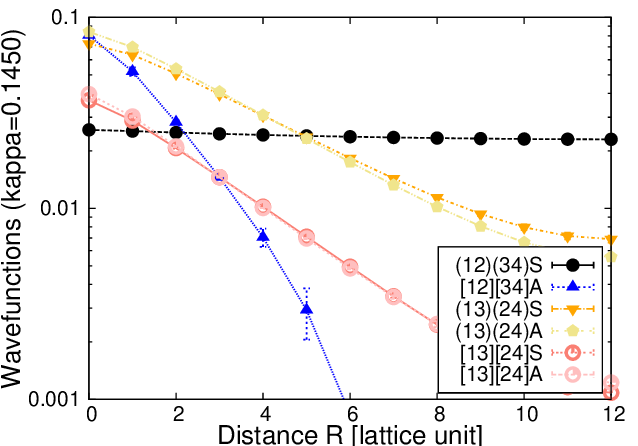}
\includegraphics[scale=0.6]{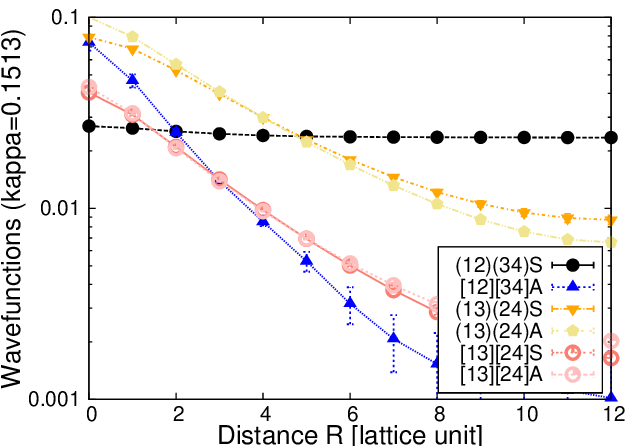}
\end{center}
\caption{
All the wavefunctions,
which are obtained without IR-gluon exchanges,
are logarithmically
plotted as functions of distance $R$ for $\kappa=$0.1450 and 0.1513.
They are normalized within $0\leq R \leq 7$.
\label{Fig.WFs2FFT}}
\end{figure}

In Fig.\ref{Fig.WFsFFT} and \ref{Fig.WFs2FFT}, the wavefunctions obtained in this framework are displayed.
1, 2-quarks and 3, 4-quarks belong to different low-energy QCD dynamics
but their high-energy dynamics are common
and high-momentum gluons can be exchanged among these four quarks.

At the source, a $(1,2)(3,4)S$-state are created
and even at the large Euclidean time only the $(1,2)(3,4)S$-state
can be observed in the asymptotic region,
which implies that asymptotic two-hadron states in other flavor channels are
successfully prohibited. 
All the other wavefunctions exponentially damp at large $R$.
On the other hand, wavefunctions for 
$[1,2][3,4]A$, $[1,3][2,4]S$ and $[1,3][2,4]A$ channels
exponentially damp toward the asymptotic region,
which remains unchanged as compared to the previous analyses.

\begin{figure}[h]
\begin{center}
\includegraphics[scale=0.6]{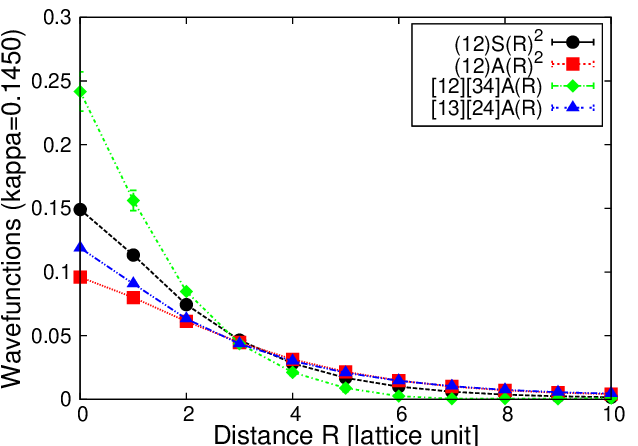}
\includegraphics[scale=0.6]{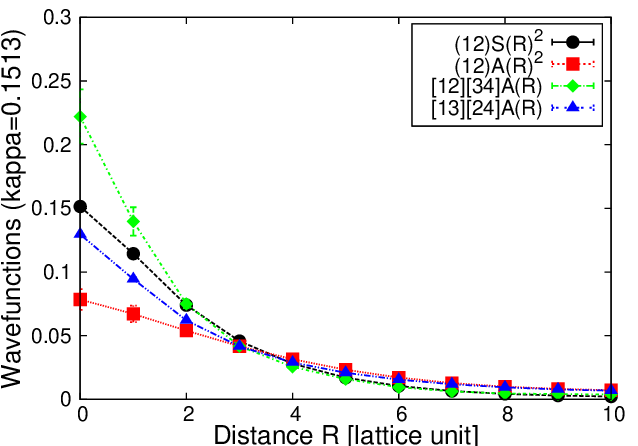}
\end{center}
\caption{
The wavefunctions measured by colored operators
as well as the squared wavefunctions of single hadrons,
are plotted.
They are obtained without IR-gluon exchanges and
are normalized within $0\leq R \leq 7$, are plotted.
\label{Fig.WFs3FFT}}
\end{figure}

\begin{figure}[h]
\begin{center}
\includegraphics[scale=0.6]{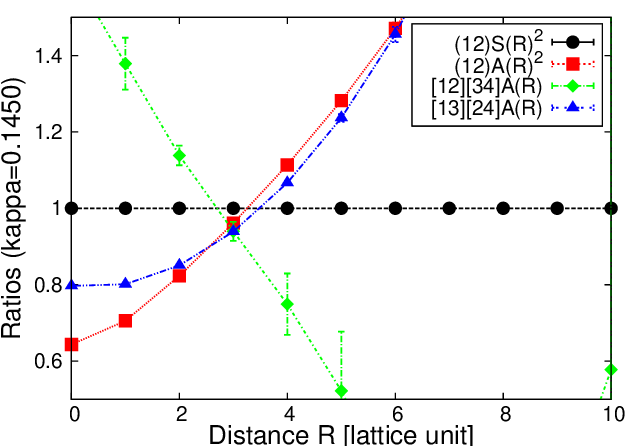}
\includegraphics[scale=0.6]{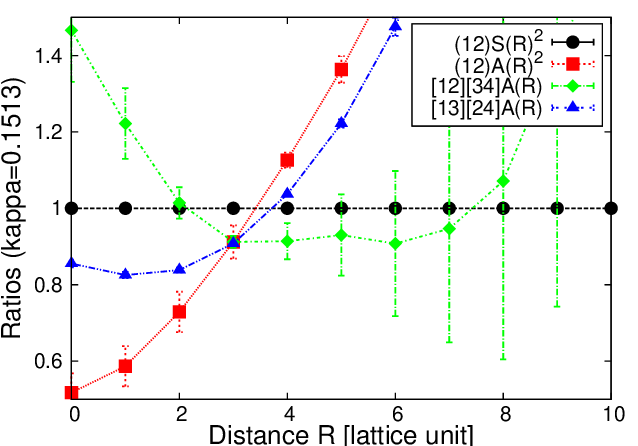}
\end{center}
\caption{
Ratios between the wavefunctions from colored operators
and the squared wavefunction of the scalar diquark hadron,
which are normalized within $0\leq R \leq 7$, are plotted.
They are obtained without IR-gluon exchanges.
\label{Fig.RTsFFT}}
\end{figure}

In Fig.~\ref{Fig.WFs3FFT} and \ref{Fig.RTsFFT},
squared internal wavefunctions, $\Phi_{(12)S}(R)^2$ and $\Phi_{(12)A}(R)^2$
for scalar- and axialvector diquarks,
and relative wavefunctions,
$\Phi_{[12][34]A}(R)$ and $\Phi_{[13][24]A}(R)$, are plotted.
All the wavefunctions are normalized within $0\leq r \leq 7$.
The colored contributions
$\Phi_{[12][34]A}(R)$
remains unchanged and is remarkably enhanced again at the short range
for the larger $\kappa$.
Now that $(13)(24)S$ and $(14)(23)S$ asymptotic scattering waves do not
exist, ``color excitation'' measured by $\Phi_{[12][34]A}(R)$
is safe from trivial colored contaminations from asymptotic scattering states.
We can conclude that
$\Phi_{[12][34]A}(R)$ around the origin mainly originates from
genuine color excitation modes.

On the other hand, $\Phi_{[13][24]A}(R)$ does not show the short-range enhancement
and lie between $\Phi_{(12)S}(R)^2$ and $\Phi_{(12)A}(R)^2$,
which implies that
$\Phi_{[13][24]A}(R)$ in the present analysis
little contains genuine color excitation components.
The reason for that is the suppression of incoming and outgoing
$(13)(24)S$ and $(14)(23)S$ scattering waves:
If $(13)(24)S$ incoming amplitude is absent,
the magnitude of color-excited $\{13\}\{24\}A$ state 
will be inevitably suppressed.
Then $\Phi_{[13][24]A}(R)$ mainly contains
trivial colored state
from the incoming and outgoing $(12)(34)S$ scattering waves,
which is not suppressed in this analysis.

\section{Discussions}
\label{Sec.Discussions}

We have found so far that
\begin{itemize}
\item
The enhancement of $\Phi_{[13][24]A}(R)$ at the short range
shows the manifestation of the color excitation modes in hadron scattering.
\item
The color excitation modes much more appear when quark masses are lighter.
\item
The color-excitation is intense at $R< 2$ in the lattice unit
($R< 0.2\  {\rm fm}$).
\end{itemize}

\begin{figure}[h]
\begin{center}
\includegraphics[scale=0.6]{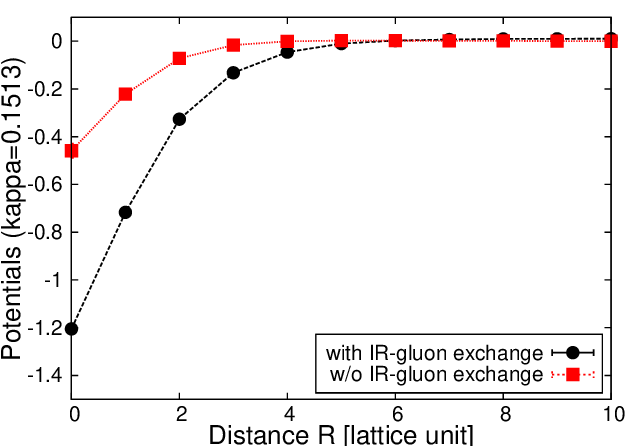}
\end{center}
\caption{
Reconstructed potentials for $(12)(34)S$ channel
at $\kappa=0.1513$.
The circles are the potential obtained by the full gluon dynamics
and 
the squares are that obtained without IR-gluon exchanges.
\label{Fig.pots}}
\end{figure}

We here discuss the physical implication of the short-range color excitations
caused by ultraviolet(UV)-gluon exchanges.
When two color-singlet hadrons gets close to each other, 
color-excited (multi-quark) states are formed exchanging UV gluons.
This color excitations eventually cause 
short-range attractive force among hadrons reported in Ref.~\cite{Takahashi:2009ef}.
To see this, we construct hadronic potentials from two-hadron wavefunctions
assuming nonrelativistic Shr{\" o}dinger equations.
The reconstructed potentials for $(12)(34)S$ scattering channel
with and without IR-gluon exchanges are shown in Fig.\ref{Fig.pots}.
The potential for $(12)(34)S$ channel
with IR-gluon exchanges is nothing but the attractive potential  for the direct diagram $V_{\rm dir}$
found in our previous work~\cite{Takahashi:2009ef}.

\begin{figure}[h]
\begin{center}
\includegraphics[scale=0.6]{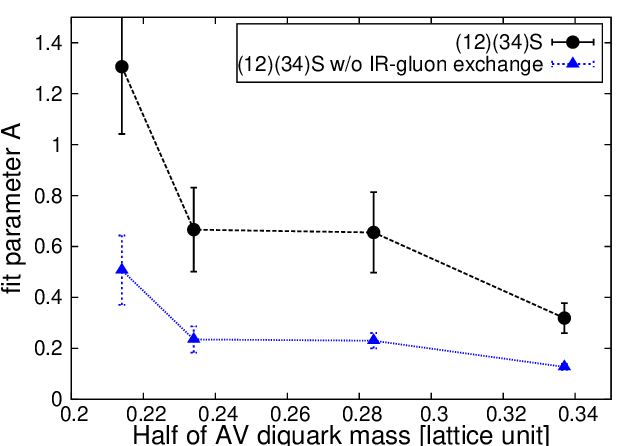}
\includegraphics[scale=0.6]{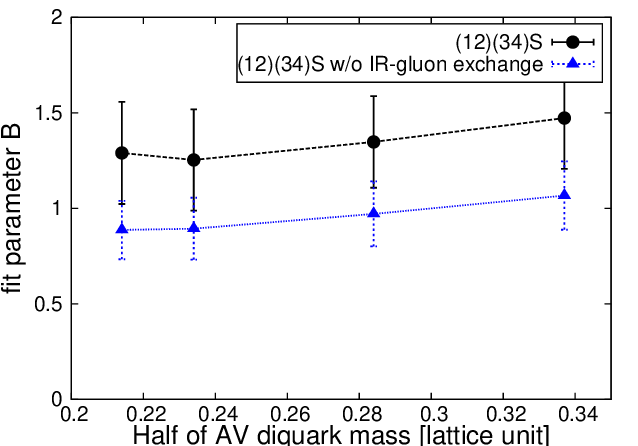}
\end{center}
\caption{Fitting parameters for the $(12)(34)S$ potentials
obtained with and without IR-gluon exchanges.
The potentials are fitted by $V=-A\exp (-Br)$.
The upper panel shows the strength $A$, and the lower panel the interaction range $B$. The horizontal axes denote the half of axial vector diquark mass
at each $\kappa$.
\label{Fig.Fittedparams}}
\end{figure}

Even when IR-gluon exchanges are made inactivated, 
an attractive potential appears between two singlet hadrons,
which can be seen as the dips around the origin.
We fit the potential with an single-exponential function
$V(R)=-A\exp (-BR)$, with $A$ and $B$ being a strength and a range
parameters.
(Such attractive potentials will include universal long-range attractive potential $V_{\rm att}^U(R)$~\cite{Takahashi:2009ef}. Since its contribution is smaller in lighter quark-mass and short-distance region, it is neglected in this analysis; $V(R)=V_{\rm dir}(R)\sim V_{\rm att}^D(R)\equiv V_{\rm dir}(R)-V_{\rm att}^U(R)$.)
We plot in Fig.~\ref{Fig.Fittedparams} the fitted parameters 
as functions of the half of the axial-vector diquark mass.
The parameters without IR-gluon exchanges are labeled as ``$(12)(34)S$ w/o IR-gluon exchange''.
Though the strength itself decreases without IR-gluon exchanges,
the interaction ranges are almost independent from quark masses
and those obtained with and without IR gluons are consistent with each other.
(When IR-gluon exchange between (12)-diquark and (34)-diquark is cut,
the energies of color combinations other than (12)(34),
including genuine 4-quark states etc., are all raised, which may have caused the reduced strength.)
The interaction range is about 1 in the lattice unit
and it is identified as the shortest-range attractive interaction $V{\rm att}$
observed in our previous work~\cite{Takahashi:2009ef}.
Taking into account that 12- and 34-diquarks do not exchange low-energy gluons
and nonperturbative interactions between 12- and 34-diquarks is absent,
this observation implies that the shortest-range gluon interaction
mainly creates colored-diquark states as intermediate states
and gives rise to the shortest-range attractive potential.

\section{Summary and Outlooks}
\label{Sec.Summary}
We have computed colored-diquark distributions in two-hadron scatterings 
using color SU(2) lattice QCD
by means of Bethe-Salpeter amplitudes on the lattice.
Using colored-diquark operators in the Coulomb gauge,
we have directly measured colored-diquark components
produced via one gluon exchange (OGE) processes in hadron scattering.
As a result, we have found that 
the colored-diquark components are much enhanced at the short range, 
$<$0.2 fm,
and that their $R$-dependence shows the tail described well by a single-exponential function.

We have repeated the analyses on the artificially constructed gauge fields,
where low- and high-momentum gluon components are decoupled and 
only restricted pair of quarks can share low-momentum gluons,
so that
we can distinguish the genuine colored-diquark components
from possible trivial colored two-quark components 
contained in two color-singlet hadrons 
as a result of 
simple transformation of hadronic basis.
Qualitatively the same behavior is actually found on this analysis,
and it is confirmed that the enhancement of the colored-diquark distribution 
at the short range we have observed 
is genuine OGE-origin color excitation.
We have also estimated the quark-mass dependence of the color-exchange range,
and the results imply that
the shortest-ranged attractive force found in two-hadron scatterings
is originated from the OGE processes between hadrons.

The short-range color-exchange processes 
produce colored many-quark objects  and enable quarks to move from one hadron to another.
One of the intriguing corresponding phenomena
may include so-called quarkyonic phase~\cite{McLerran:2007qj,Abuki:2008nm,Miura:2008gd,McLerran:2008ua,Brauner:2009gu,Hands:2010gd,Fukushima:2015bda}.
In a cold high density matter, 
quark's degrees of freedom are expected to be manifested 
at a certain density despite of its color-confining nature~\cite{Baym:1976yu,McLerran:2007qj,Abuki:2008nm,Miura:2008gd,McLerran:2008ua,Brauner:2009gu,Hands:2010gd,Fukushima:2015bda}.
In such a system, hadrons are closely located each other
and the enhanced short-range OGE processes would make hadrons colored.
This color percolation
enables quarks to move from one to another.
To cast light on such phenomena from the view point of quarks and gluons,
the much more clarification of color exchanges among hadrons at high density 
can be one of the important keys.

\acknowledgments
All the numerical calculations were performed on NEC SX-8R at CMC, Osaka university, on SX-8 at YITP, Kyoto University. This work was supported in part by the Yukawa International Program for Quark-Hadron Sciences (YIPQS), and the Grant-in-Aid for the Global COE 
Program ``The Next Generation of Physics, 
Spun from Universality and Emergence'' from MEXT of Japan,
and by KAKENHI (21740181, 25247036, 26400270).

\end{document}